\documentclass[conference]{IEEEtran}

\usepackage{tcolorbox}
\usepackage[T1]{fontenc}

\usepackage{tabularx}
\usepackage{enumitem}
\usepackage{adjustbox}
\usepackage{makecell}
\usepackage{subfig}
\usepackage{multirow}
\usepackage{hyperref}
\usepackage{booktabs}
\usepackage{balance}
\usepackage{mathtools}
\usepackage{cases}
\usepackage[absolute,showboxes]{textpos}

\usepackage{pifont}

\usepackage{xspace}

\newcommand{\etal}{\textit{et al.}}
\newcommand{\eg}{\textit{e.g.,}~}
\newcommand{\ie}{\textit{i.e.,}~}

\newcommand{\one}{({\em i})\xspace}
\newcommand{\two}{({\em ii})\xspace}
\newcommand{\three}{({\em iii})\xspace}
\newcommand{\four}{({\em iv})\xspace}

\newcommand{\loraicn}{LoRa-ICN\xspace}

\let\orgautoref\autoref
\renewcommand{\autoref}
{\def\sectionautorefname{Section}%
\def\subsectionautorefname{Section}%
\def\subsubsectionautorefname{Section}%
\orgautoref}

\makeatletter
\renewcommand{\paragraph}[1]{\vspace*{0.03in}\noindent{\bf #1.}\hspace{0.25ex \@plus1ex \@minus.2ex}}
\newcommand{\paragraphS}[1]{\vspace*{0.03in}\noindent{\bf #1}\hspace{0.25ex \@plus1ex \@minus.2ex}}
\makeatother

\begin{document}

\title{Long-Range ICN for the IoT:\\Exploring a LoRa System Design}

\author{
\IEEEauthorblockN{Peter Kietzmann\IEEEauthorrefmark{1},
Jos{\'e} Alamos\IEEEauthorrefmark{1}\IEEEauthorrefmark{3},
Dirk Kutscher\IEEEauthorrefmark{2},
Thomas C. Schmidt\IEEEauthorrefmark{1},
Matthias W\"ahlisch\IEEEauthorrefmark{3}
}

\IEEEauthorblockA{HAW Hamburg, Germany\IEEEauthorrefmark{1}
\quad
	Hochschule Emden/Leer, Germany\IEEEauthorrefmark{2}
\quad
	Freie Universit\"at Berlin, Germany\IEEEauthorrefmark{3}
}

\IEEEauthorblockA{\{first.last\}@\{haw-hamburg.de, hs-emden-leer.de, fu-berlin.de\},
t.schmidt@haw-hamburg.de}
}

\IEEEoverridecommandlockouts
\IEEEpubid{
   \makebox[\columnwidth]{ISBN 978-3-903176-48-5~\copyright~2022 IFIP \hfill}
   \hspace{\columnsep}
   \makebox[\columnwidth]{ }
}

\maketitle

\setlength{\TPHorizModule}{\textwidth}
\setlength{\TPVertModule}{\paperheight}
\TPMargin{5pt}
\begin{textblock}{1}(.1,0.02)
  \noindent
  \footnotesize
  If you cite this paper, please use the IFIP Networking reference:
  P. Kietzmann, J. Alamos, D. Kutscher, T. C. Schmidt, M. W\"ahlisch.\\
  Long-Range ICN for the IoT: Exploring a LoRa System Design.
  \emph{Proc. of 21th IFIP Networking Conference}, IEEE, 2022.
\end{textblock}

\begin{abstract}
    This paper presents LoRa-ICN, a comprehensive IoT networking system based
    on a common long-range communication layer (LoRa) combined with
    Information-Centric Networking~(ICN) principles.  We have replaced the
    LoRaWAN MAC layer with an IEEE 802.15.4 Deterministic and Synchronous
    Multi-Channel Extension (DSME). This multifaceted  MAC layer allows for
    different mappings of ICN message semantics, which we explore to enable
    new LoRa scenarios.

    We designed LoRa-ICN  from the ground-up to improve reliability
    and to reduce dependency on centralized components in
    LoRa IoT scenarios.
    We have implemented a feature-complete prototype in a common network
    simulator to validate our approach. Our results show design trade-offs
    of different mapping alternatives in terms of robustness and~efficiency.
\end{abstract}


\section{Introduction}\label{sec:intro}

LoRaWAN is a popular low-power long-range communication system for IoT
that is suitable for single-site deployments as well as for larger
networks.
It consists of LoRa, a PHY layer that allows for radio communication between 2 and 14~km, and higher-layer protocols mainly to upload IoT data to a server-based infrastructure.
These characteristics make LoRaWAN a promising option for many urban and rural IoT scenarios.

The LoRaWAN network design incurs, however, four notable shortcomings:
\one~LoRaWAN is heavily optimized towards retrieving data \emph{from} constrained
    Nodes.  Sending data \emph{to} Nodes is expensive and involves
    significant latencies. Many networks such as the popular community {\em The Things Network}~(TTN) thus deprecate sending
    data to Nodes above a very low message rate, making LoRaWAN unsuitable
    for most control scenarios.
\two~LoRaWAN has not been designed with the objective to provide a platform for Internet protocols.
    It is possible to use IP and adaptation layers
    on top of LoRaWAN, albeit very inefficiently.
\three~The whole LoRaWAN system is a vertically integrated stack
    that leads to inflexible system
    designs and inefficiencies. For example, all
    communication is channeled through  LoRaWAN Gateways
    as well as Application- and Network Servers that interconnect with
    applications. 
\four~The centralization and lock-in into vertical
    protocol stacks challenge data sharing (between users)
    and the creation of distributed applications (across LoRa island and the~Internet).

In this paper, we
aim for a better integration of the LoRa-based Internet of Things into the remaining
Internet.
We base our system design on the following four
requirements:
\one~enabling LoRa networks and Nodes in these networks to communicate
directly with hosts on the Internet;
\two~empowering LoRa Gateways to act as routers, without the need to
employ Network Servers and to tunnel all traffic to or from them;
\three~enabling secure data sharing and wireless Node control;
\four~maintaining the important power conservation and robustness
properties of current LoRaWAN systems.

To achieve these goals without abandoning the benefits of the LoRA PHY
(\ie a robust, energy-efficient long-range communication channel) we
propose both a complete redesign of the MAC layer and a data-driven layer on top.
Our proposal leverages two key building blocks.  First, the Deterministic
and Synchronous Multi-Channel Extension (DSME) extension to IEEE
802.15.4e~\cite{IEEE-802.15.4-16}, a flexible MAC layer that consists
of contention-access and contention-free periods, and, second, the
Infor\-mation-Centric Networking (ICN) protocol NDN~\cite{zabjc-ndn-14}, which
provides secure access to named data in networks.

Prior work showed that ICN provides clear benefits  over traditional IP and CoAP or MQTT stacks in the IoT~\cite{gklp-ncmcm-18}, its integration into LoRa is missing, yet.
We argue that ICN  is well-suited for use with LoRa because its hop-wise data replication  increases robustness and flexibility while reducing retransmission load.  This enhances adaptivity and decreases communication overhead, whereas link capacity is scarce with LoRa. Named and authenticated data access enables location-independence since applications can access named data directly, without resorting to lower-layer addresses.
Furthermore, built-in caches in ICN facilitate more efficient LoRa networks. Requests that are satisfied by an in-network cache
\one~reduce link utilization, to improve on air time and wireless interference,
\two facilitate Node sleep, and
\three reduce long round trips introduced by slow transmissions.

\noindent In summary, our main contributions are:

\begin{enumerate}[wide, labelwidth=!, labelindent=0pt,itemsep=3.6pt]
    \item The design of ICN over LoRa, including a suitable DSME
    configuration and options for mapping ICN messages to DSME.  (\textbf{\S~\ref{sec:design-goals}} and
    \textbf{\S~\ref{sec:concept}})
    \item A complete simulation environment in OMNeT++ that combines ccnSim as an ICN stack, openDSME as a MAC layer, and FLoRa to simulate LoRa-type devices---and a demonstration of our adaptation layers in that system.
      (\textbf{\S~\ref{sec:impl}})
      \item Based on our simulation results, we derive preferred
        mappings and additional Node requirements for implementing
        relevant ICN interaction patterns.
        (\textbf{\S~\ref{sec:eval}} and \textbf{\S~\ref{sec:design-conclusions}})
\end{enumerate}

\section{Background and Challenges}
\label{sec:ps}

\paragraph{LoRa PHY: Long-range but very low data rates}
The LoRa PHY layer defines a chirp spread spectrum modulation
which enables a long transmission range (theoretical
2--14\,km) using minimal energy. Spreading
factor (SF), code rate, and bandwidth can be configured and directly
affect the time on air and data rate. As an example, a 50\,Bytes frame
has an on-air time of
2.3\,seconds using SF12, code rate 4/5, 125\,kHz
bandwidth which leads to a PHY bit rate of 250\,bit/s.
Varying center-frequencies in
the sub-GHz ISM band constrain the duty cycle to 0.1--10\,\% and
further limit the effective throughput. As a consequence, the maximum
effective bitrate of the physical layer can be as low as
0.25\,bit/s.

While the LoRa PHY provides attractive features, it clearly imposes
significant constraints with respect to worst-case latency and
throughput, regardless of higher layer protocols such as the MAC
layer. It is important to note that LoRa networks are therefore not
comparable to IEEE 802.11---instead they provide properties
that incur significant challenges to higher-layer protocol design with
respect to delay tolerance.

\paragraph{LoRaWAN MAC Layer: Limited communication models}
The LoRaWAN MAC layer defines three operation modes:
classes~A--C.  In class~A, constrained Nodes send uplink using the
ALOHA medium access protocol and can receive downlink traffic within
two subsequent slots. This approach has three limitations: \one ALOHA
is susceptible to collisions.  \two Downlink traffic is fairly limited
and cannot be initiated by a Gateway.  \three It prevents broadcast
traffic.  Class~C works similarly to class~A but leaves the radio
always on, which enables Gateway initiated (multicast) downlink
traffic but comes at high energy cost on constrained Nodes.  Class~B
adds periodic slots which allows for ``predictable'' downlinks at
medium energy consumption.  Gateways send beacons every 128\,seconds
(time-synchronized by GPS). Consequently, hardware requirements are
not compatible with current deployments that mostly serve
class~A. Beacons of neighbored Gateways can collide, for the absence
of beacon synchronization.  The Network Server arranges
MAC schedules, however, downlink slots can overlap or be
suspended. Scalability issues of class~B have been analyzed
in~\cite{sak-lcbmc-20}.  Furthermore, uplink traffic
still uses ALOHA in class~B mode, which interferes with downlink
traffic, so that communication is still best effort.

\paragraph{LoRaWAN Network Architecture: Gateway and server centric}
In the LoRaWAN architecture, radio networks are connected by
Gateways to a Network Server that provides over-the-air activation,
message deduplication, message routing, adaptive rate control at end
devices, and acknowledging messages. Gateways are merely relays that
implement timing-relevant aspects of the MAC protocol such as sending
beacons.
In the upstream direction, Gateways forward (tunnel) frames to the Network
Server over the Internet.
In the downstream direction, the Network Server sends LoRaWAN messages
to LoRa Nodes, which includes Gateways selection.

\begin{figure}
  \subfloat[Layer configuration.]{%
     \scalebox{0.7}{\includegraphics{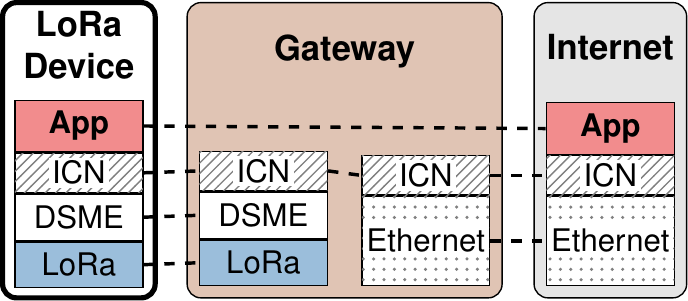}}
     \label{fig:stacks}
  }
  \subfloat[Network example.]{%
     \scalebox{0.7}{\includegraphics{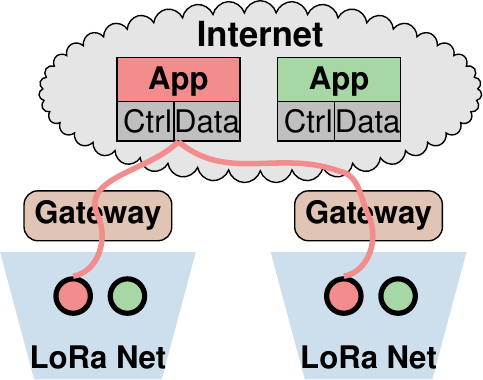}}
     \label{fig:network}
  }
  \caption{\loraicn stacks and networks.}
  \label{fig:stacksandnetwork}
\end{figure}

The LoRaWAN specification does not mandate particular deployment
options, and Network Servers could in theory be co-located to
Gateways. In practice, \eg
in public networks, such as TTN~\cite{ttn}, Network
Servers are operated in the Internet, and
Gateways are peripherals of the Network Server, \ie they cannot
operate without it.  While this design decision is practical with
respect to ease Gateway operation, it leads to a centralized
architecture around the Network Server and additional servers such as
Application Servers that provide interfacing to business
logic, interconnecting local LoRa networks, and data sharing.

\paragraph{DSME: Reliable and ICN-friendly MAC layer}
The 802.15.4 DSME MAC (see~\cite{IEEE-802.15.4-16} for further details) 
 enables new LoRa scenarios.
A coordinator starts network formation and emits beacons in a pre-defined beacon interval (including beacon collision resolution mechanisms), to initiate a synchronized multi superframe structure. A superframe in the multi superframe consists of: Beacon period (BP), contention-access period (CAP), and contention-free period (CFP); the latter of which provides seven guaranteed time slots (GTS), multiplexed across 16 radio channels.
Data is transmitted during the CAP, a pre-allocated slot in the CFP, or in an ``overloaded'' beacon.
Battery-driven Nodes mostly sleep (\eg during the CAP), which makes them unavailable for the coordinator.
DSME provides \textit{Indirect Transmission}, in which a coordinator indicates pending transactions with a beacon. This triggers the constrained Node to stay awake after the BP.

\section{Design Goals}\label{sec:design-goals}

In \loraicn, we are re-imagining the system architecture of long-range
IoT communications, aiming to overcome the challenges described in the
previous section.  In our deployment scenarios, an application
consists of the control and data consumer applications in the Internet
and a set of \loraicn Nodes (ICN producers and consumers), potentially
distributed over multiple individual LoRa radio networks.  Each LoRa
network is served by one Gateway.  The controlling and consuming
applications can access all their associated wireless Nodes directly
over the Internet, without mediation through application layer
Gateways, see Figure~\ref{fig:stacksandnetwork}.

Our design goals are \one not requiring changes to the rest of the ICN
network, \two providing a complete set of interaction patterns such as
data transmission and Node control, 
\three~leveraging the LoRa PHY capabilities optimally.  To implement a
fully distributed system model, \loraicn Gateways operate as layer~3
routers instead of just bridges as in LoRaWAN. Key functions that are
typically implemented by LoRaWAN Network Servers (\eg routing) are
performed by Gateways.

In order to leverage the LoRa PHY capabilities and to
support the rich ICN interaction patterns, we replace the LoRaWAN MAC
layer with the significantly more powerful IEEE 802.15.4 DSME MAC layer~\cite{IEEE-802.15.4-16} that provides better
reliability (important for ICN Interests) and reduced latency
(important for ICN consumer-publisher communication).
In the following, we discuss the most relevant operations that support our deployment scenarios.

\paragraphS{Node Registration} refers to a Node registering its prefixes with
the local Gateway (acting as ICN forwarders), which will install FIB
entries and announce the prefixes outside the LoRa network so that Nodes
do not need to participate in routing.

\paragraphS{Data Provisioning by Nodes} on LoRa Nodes includes asynchronously
produced sensor data as well as requested data transmission. ICN is a
receiver-driven system, so we distinguish two main variants: \one
\emph{ICN-idiomatic Interest/Data}.
\two \emph{Push Data from Nodes to Gateways}. While unsolicited push
is not an ICN-idiomatic communication pattern, it is still a useful
capability in a resource-constrained environment because IoT Nodes
may need to save energy and produce data only occasionally.

\paragraphS{Node Control} refers to Nodes being reliably controlled by peers
in the Internet, \eg for sensor control and configuration. We use a
basic Interest-triggered interaction, and Data as ACK.

\paragraphS{Data Retrieval by Nodes} is natively enabled as
\loraicn Nodes are regular ICN Nodes and may also send Interests to
other ICN Nodes hosted by the same LoRa network, other LoRa networks, or 
any other ICN network.

\paragraphS{Downstream Multicast} enables large-scale data distribution as needed for example in firmware updates. ICN features multicast
via Node-generated Interests and broadcast Data
messages from the Gateway.

\section{ICN over LoRa}\label{sec:concept}

The LoRa PHY exhibits long on-air times (seconds) for transmit long-
range (kilometers) at minimum energy consumption (micro-joules) and
underlies rigorous duty-cycle restrictions. In order to achieve a robust system design, we proceed in two steps. First, we utilize a proven LoRa PHY configuration to leverage the DSME MAC. Next, we derive a
viable mapping of ICN to DSME/LoRa for sending ICN Interest and Data messages from and to LoRa Nodes.

\subsection{Mapping DSME to LoRa}\label{sec:concept_dsmelora}

We apply the PHY mapping presented by Alamos~\etal~\cite{aksw-dfml-21} to utilize LoRa below DSME.
This includes
a spreading factor of 7,
a bandwidth of 125\,kHz,
and a code rate of 4/5.
The beacon interval is 125.82\,s to align with LoRaWAN class B beacons (128\,s).
The contention-free period (CFP) defines 16 channels with a 1\,\% duty-cycle restriction. Beacons and CAP use a common channel of 10\,\% duty cycle. During CAP, Nodes perform CSMA-CA and incorporate channel activity detection (CAD) of common LoRa devices.

The CFP channels are designed to carry data of high reliability and limited latency demands.
The time division of DSME, though,  requires a packet queue and hence affects transmission speed. Traffic load determines queue occupation.
We want to estimate the average waiting time (\ie time in queue) for a packet that should be transmitted reliably during the CFP.

Little's law~\cite{l-pqf-61} $W=L/\lambda$ approximates the average waiting time $W$, using the average number of items $L$ (\ie queued packets) at a given average arrival rate $\lambda$ (packet rate).
We assume \one independent, exponentially distributed inter arrival times of packets with an
expectation rate of $\lambda$. \two Nodes allocate only one transmission slot in CFP and (without loss of generality) \three the MAC queue has infinite capacity.

Let $L(t_n)$ be the number of queued packets after the transmission time $t_n$ of the \emph{n}-th multi superframe. We note that $(L(t_n))_n$ is an ergodic Markov process (positive recurrent and aperiodic), for which the limiting distribution $\pi_i$ exists (with $i= 0,\ldots,\infty$ the number queued packets). This stationary eigenvector can be calculated numerically as a fixed point of a (clipped) high-dimensional transition matrix and yields the stationary mean occupancy  $L(t_\infty)$ prior to starting the new superframe. The actual  queue that an arriving packet faces holds also packets which arrived during the current multi superframe (of duration $T$), \ie

\begin{equation}
L = L(t_\infty) + \frac{\lambda \cdot T}{2}
\end{equation}

As an example, we choose a relaxed packet arrival rate  $\lambda=1/120$\,s and compose our multi-frame structure of $4$ superframes, \ie $T =32.46$\.s. This results in an average number of $L \approx 0.18$ queued packets and an average waiting time of $W \approx 21.32$\,s respectively. This scenario is compatible to the downlink in a common class B configuration, which exhibits an average waiting time of $44$\,s but suffers from  26\% loss~\cite{ekbb-elcbe-20}. In contrast, loss is very unlikely in our CFP time division period.

\begin{figure*}
    \centering
    \scalebox{.8}{\includegraphics{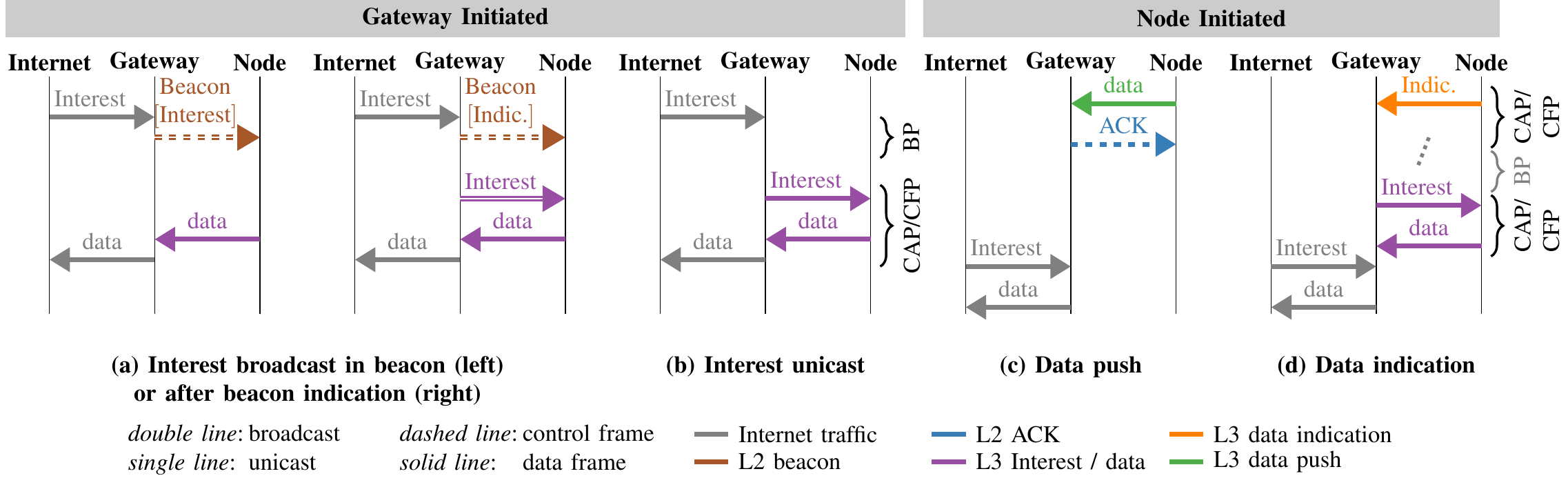}}
    \caption{Mapping schemes of Interest/Data and ICN extension packets to
    DSME frames. Data flows from Node to Gateway, either initiated by the
    Gateway ((a)-(b)) or by the Node ((c)-(d)).}
    \label{fig:sequences3}
\end{figure*}

\subsection{A MAC for ICN using a LoRa-Proxy}\label{sec:concept_icndsme}

\autoref{fig:sequences3} presents options to handle Interest and Data
packets between a high throughput network (\eg the Internet) and a
DSME/LoRa network.  The left part (\autoref{fig:sequences3}(a)-(b)) shows Gateway-initiated
request-response communication, resembling native ICN primitives.
In addition, we present protocol extensions for Nodes to initiate traffic (\autoref{fig:sequences3}(c)-(d)) to
Gateways that act as proxies for constrained Nodes during sleep time.

\paragraph{LoRa$\rightarrow$Internet. Broadcast}
Beacons are regularly broadcast by Gateways and can carry Interests
without message overhead. This maximizes sleep cycles, but the limited
beacon intervals reduce throughput.
Using beacons to transfer Interests provides two options (see Fig. \ref{fig:sequences3}(a)).
\one beacons carry payloads up to a frame size of  127\,Byte minus metadata, \ie  $\approx$ 100\,Bytes/frame.
Using ICNLoWPAN encoding~\cite{gksw-innlp-19} this is sufficient to aggregate 4--6 Interest packets.
\two Gateways utilize indirect transmission (see~\autoref{sec:ps}) to broadcast an Interest,
which involves the indication of a pending transaction within the
beacon, and subsequent Interest broadcast during the CAP.

\paragraph{LoRa$\rightarrow$Internet. Unicast}
Interest and Data messages can be sent via unicast within the CAP or
CFP (see Fig. \ref{fig:sequences3}(b)).  Sending Interests in best effort CAP frames enables requests at
higher rates than in beacons. Nevertheless, this prevents Nodes from
sleeping during the CAP.
Note that individual Interests increase the number of downlink packets
from the Gateway; for growing wireless networks this conflicts with
duty cycle restrictions at the Gateway.  Using the CAP for Data
instead is less critical since transmissions are initiated in the
low-power Node. The CFP provides exclusive resource access but adds
the overhead of a preceding cell negotiation, and is limited within the
superframe structure.  Hence, a full CFP Interest-Data exchange
requires two cell allocations per Gateway-Node pair.

\paragraph{LoRa$\rightarrow$Internet. Data indication}
 Nodes can offload the Gateway by removing the need for polling. This is done
 indicating names for subsequent Interests from the Gateway  (see Fig. \ref{fig:sequences3}(d)).
The indication packet, however, adds wireless traffic.
Since Nodes do not emit beacons, indication packets utilize unicast
traffic in the CAP or CFP. Interest and Data packets follow as
outlined in Figs. \ref{fig:sequences3}(a) or (b). Hence, an indication
and the following Interest broadcast must wait for the next beacon
period. Instead, Interest unicast in CAP or CFP keeps the latency for
producer-initiated traffic minimal.

\paragraph{LoRa$\rightarrow$Internet. Local Data push}
A link-local Data push from producers to the Gateway reduces radio
access and maximizes device sleep times, similar to LoRaWAN~class~A
deployments (see Fig. \ref{fig:sequences3}(c)).  An optional link layer ACK with retransmissions from the
Node increases reliability in the CAP; the exclusive CFP slots can
omit the ACK.

\begin{figure}
    \centering
    \scalebox{.8}{\includegraphics{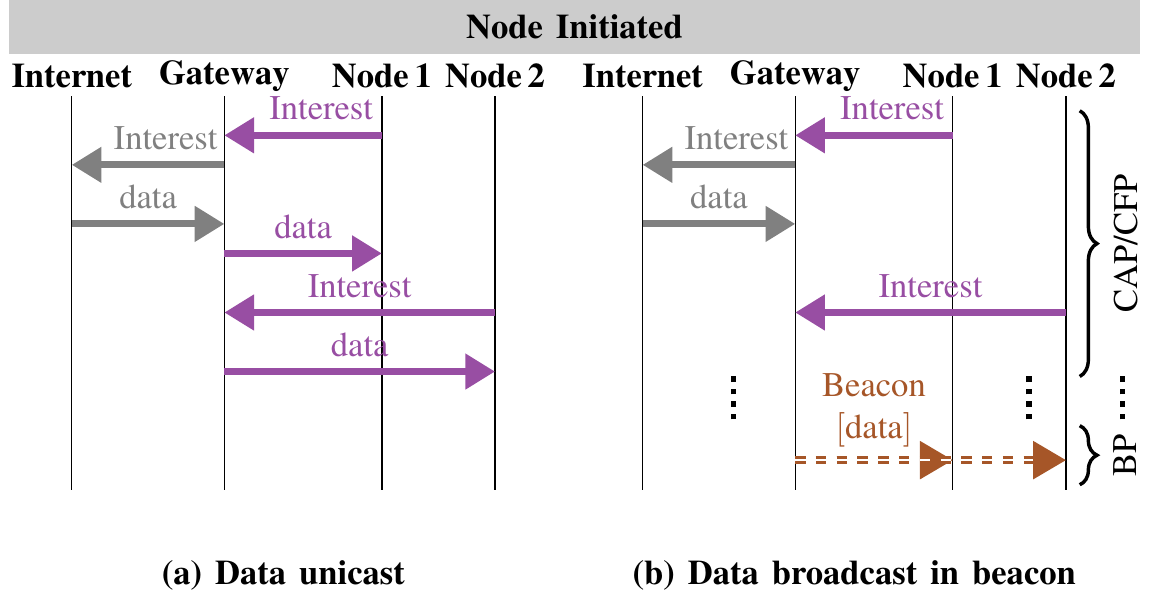}}
    \caption{Mapping schemes of Interest/Data and ICN extensions packets to
    DSME frames. Data flow is from Gateway to Node.}
    \label{fig:sequences_nodereq}
\end{figure}

\paragraph{Internet$\rightarrow$LoRa. Unicast}
Nodes request data from a Gateway by sending Interests within CAP or
CFP frames (\autoref{fig:sequences_nodereq} (a)).  Data returns in a
CFP cell pre-allocated for every consumer Node.  This provides high
reliability but becomes challenging in larger networks, since the amount
of downlink cells is limited by the multi superframe.

\paragraph{Internet$\rightarrow$LoRa. Broadcast}
Multiple Interests for the same content item arrive at the Gateway
that responds with a Data broadcast message (see
\autoref{fig:sequences_nodereq}(b)) using an overloaded beacon. This
can help with observing duty cycle restrictions on the Gateway;
however the data throughput is limited by the beacon interval.

\section{Simulation Environment}\label{sec:impl}

\begin{figure}
    \centering
    \scalebox{.825}{\includegraphics{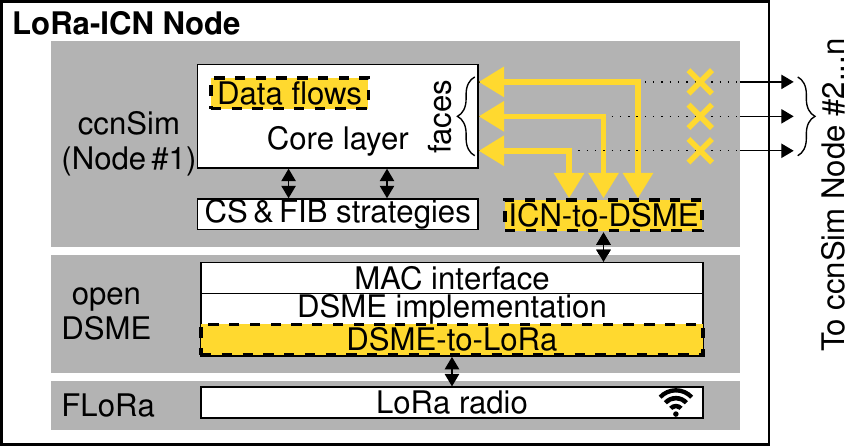}}
    \caption{Simulation environment and our extensions.}
    \label{fig:overview_simulation}
\end{figure}

We have developed a simulation environment for \loraicn that is based
on OMNeT++ and the INET
framework~\cite{inet-framework-21}. We integrated
ccnSim~\cite{crr-chscs-13} for core ICN support and
openDSME \cite{kkt-rwmnd-18} for 802.15.4 DSME
functionality. Our model uses FLoRa~\cite{spd-aclnd-18} and its wireless propagation model and PHY.
Figure \ref{fig:overview_simulation} depicts the simulation environment and
our extensions.

{\bf Data flows} orchestrate ICN  Interest/Data exchanges and are
adjusted to match IoT use cases as follows: We changed the build-in
content popularity model from Mandelbrodt-Zipf to a uniform
distribution and initiate one transmission for every content item.  We
also extended the ccnSim core implementation by two network layer
primitives: \one \textit{Indication}
(see~\autoref{sec:concept_icndsme}) and  \two \textit{Push} to place a data item in the neighboring content store.
In all scenarios, content rates follow a Poisson process.

\paragraphS{ICN-to-DSME} addresses three main challenges.
\one ccnSim lacks the concept of a link layer. Instead, ICN faces
directly connect to I/Os of neighboring Nodes. We
 include a wireless transmission link.
\two We add a face-to-MAC module that multiplexes ICN faces to a
\textit{WirelessInterface} and uses MAC addresses for
transmission.  This module includes all logic for the ICN-to-DSME mapping
(see~\autoref{sec:concept_icndsme}).
\three OMNeT++ messages are converted into INET packets that map to openDSME. This step includes
tagging of packets and appends control instructions for the MAC
layer.

\paragraphS{DSME-to-LoRa} integrates the LoRa PHY with
the DSME MAC implementation. This component bases on related work and we refer the reader to Alamos~\etal~\cite{aksw-dfml-21}.
We further disable dynamic slot allocation for the CFP to exclude
negotiation overhead, but implement static scheduling and MAC
configurations. In bidirectional communications, each Tx slot is
followed by an Rx slot which halves the number of transactions in one
multi superframe. Every simulation
Node is assigned zero, one, or two slots depending on the MAC
mapping. This pattern repeats with a multi superframe -- with adjustable structure to simulate different network sizes.

\section{Evaluation}\label{sec:eval}

We have implemented and tested different options for ICN-to-DSME
mappings for the two major use cases {\em Data from Node to Gateway}
and {\em Data from Gateway to Node}.

\subsection{Data from Node to Gateway}\label{sec:data2gw}

\begin{table}[b]
    \centering
    \caption{Performance overview of mapping schemes.}
    \label{tbl:gw_client}
    \begin{adjustbox}{max width=1\columnwidth}
        \setlength{\tabcolsep}{3pt}
        \begin{tabular}{l  l r  r  r  r  r r}
            \toprule
            &Mapping Scheme &Indication&Interest&Data& \makecell[c]{Avg.\\Latency\,\\$[$s$]$}&\makecell[c]{Max.\\Latency\,\\$[$s$]$} & \makecell[c]{Data\\Loss\\$[$\%$]$}\\
            \midrule
            \multirow{6}{*}{\rotatebox[origin=c]{90}{\makecell[l]{\colorbox{lightgray}{Gateway Initiated}}}}
            &\multirow{2}{*}{\rotatebox[origin=c]{00}{\makecell[l]{Interest broadcast}}}
            &&Beacon&CAP&\multicolumn{3}{c}{\multirow{2}{*}{(Not operable at this scale)}}\\
            &&&Beacon&CFP&&&\\
            \cmidrule(lr){2-8}
            &\multirow{2}{*}{\rotatebox[origin=c]{00}{\makecell[l]{Interest unicast\textsuperscript{1}}}}
            &&CAP&CAP& 8.32&56.56&5.16\\ 
            &&&CAP&CFP& 11.67&26.87&0.05\\ 
            \cmidrule(lr){2-8}
            &\multirow{2}{*}{\rotatebox[origin=c]{00}{\makecell[l]{Interest unicast}}}
            & &CFP&CAP& 12.73 & 28.27&1.72\\ 
            && &CFP&CFP& 17.01 & 47.24 &0.00\\ 
            \midrule
            \multirow{6}{*}{\rotatebox[origin=c]{90}{\makecell[l]{\colorbox{lightgray}{Node Initiated}}}}&\multirow{4}{*}{\rotatebox[origin=c]{00}{\makecell[l]{Data indication}}}
            &CAP&CFP&CAP& 19.25&75.17&3.13\\ 
            &&CAP&CFP&CFP& 14.61&71.61&1.51\\ 
            &&CFP&CFP&CAP& 71.62&299.22&2.6\\ 
            &&CFP&CFP&CFP& 46.49&241.96&0.89\\ 
            \cmidrule(lr){2-8}
            &\multirow{2}{*}{\rotatebox[origin=c]{00}{\makecell[l]{Data push}}}
            & &&CAP& 7.02&29.88&1.62\\ 
            && &&CFP& 10.63&66.42&0.00\\ 
            \bottomrule
        \end{tabular}
        \smallskip
    \end{adjustbox}
    \begin{minipage}{\textwidth}
       {\footnotesize\textsuperscript{1}Nodes must be turned on during CAP, which prevents low-power.}
    \end{minipage}
\end{table}

\paragraphS{Motivation:}
\autoref{tbl:gw_client} presents an overview of the performance for
ICN/DSME/LoRa in a network of 14 Nodes, when the Node is a producer,
and data flows towards the Gateway.  For {\bf Gateway-initiated} traffic,
{\bf Interest broadcast} reflects a special case which is heavily
limited by the beacon interval. With an aggregation of multiple
Interests into one broadcast message, the Gateway is able to send
$\approx$ 5 Interests encoded in one broadcast packet, every $\approx$ 126\,s (beacon interval).
Consequently, we can accommodate up to 5 Nodes responding with
one Data message each in the same multi superframe.

\begin{figure*}
    \centering
        \subfloat[Content at 30\,s interval per Node]{\includegraphics[scale=0.9]{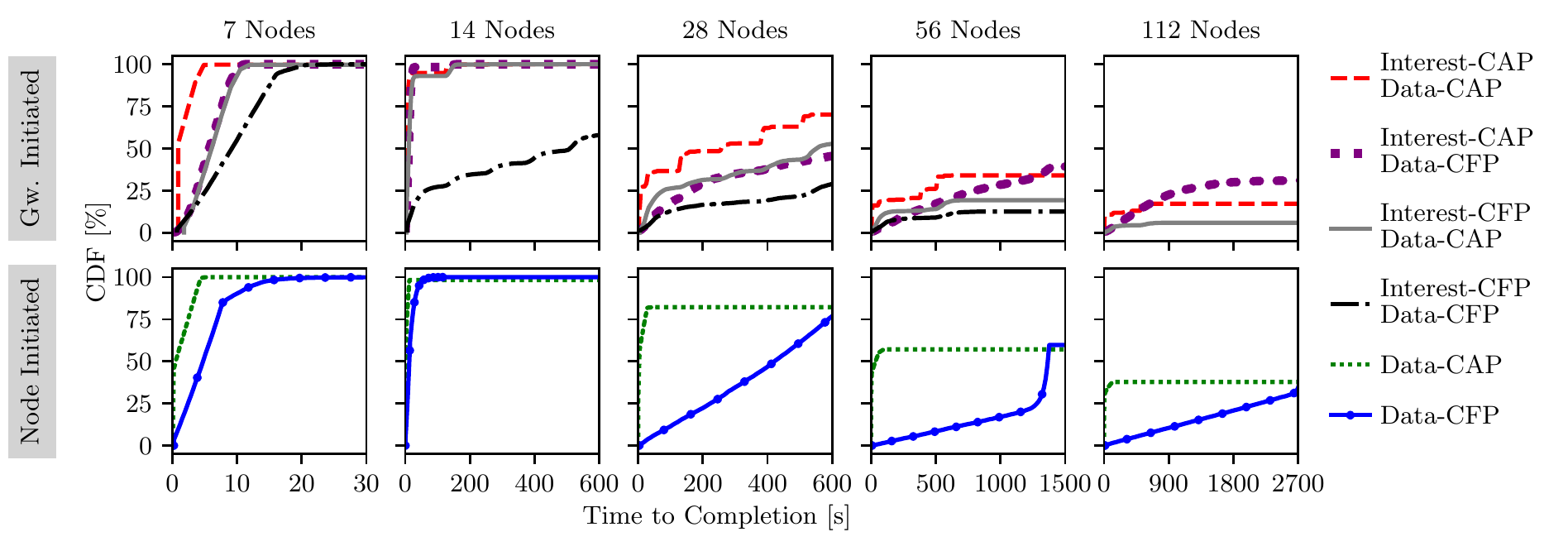}\label{fig:ttc_30_exponential_4}}\hfill
        \subfloat[Content at 900\,s interval per Node]{\includegraphics[scale=0.9]{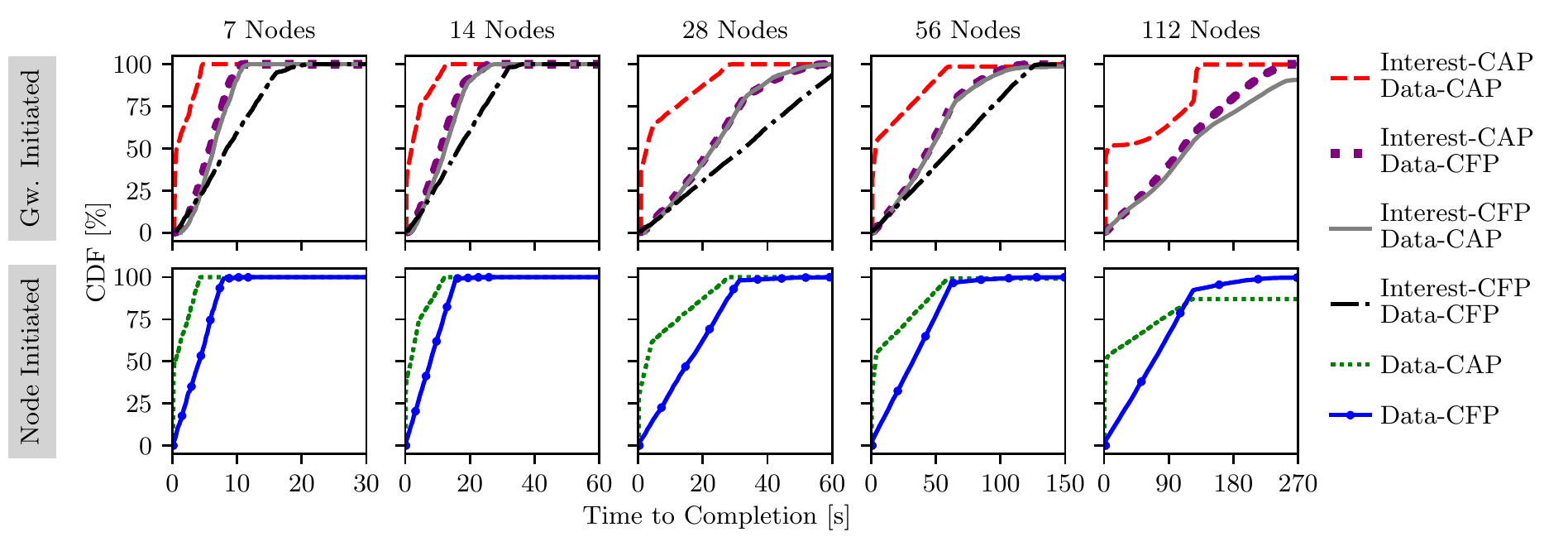}\label{fig:ttc_900_exponential_4}}
    \caption{Time to content arrival with different mapping schemes for Interest/Data (with L3 retransmission) and Data push for varying network sizes.}
    \label{fig:ttc_exponential_4}
\end{figure*}

For the other mappings (see~\autoref{sec:concept_icndsme})
in \autoref{tbl:gw_client}, each of the 14 Nodes produces a content item
in one minute intervals (on average), and we disable Interest
retransmissions to evaluate the plain ICN performance over LoRa.  {\bf
  Interest unicast} is separated into {\bf Interest-CAP and CFP}
variants (Interest-CAP prevents Nodes from sleep and is not feasible
for battery powered devices).  {\bf Interest-CAP} provides short
completion times of 8--12\,s on average, due to frequent CAP
intervals. {\bf Interest-CFP} is slower by a factor of $\approx$ 1.5.
However, sending data in the CAP increases the probability for data
loss, which is most notable when Interest and Data messages share the
CAP. The maximum latency increases up to 56\,s as a consequence of
CSMA retries. Conversely, sending Data messages in the CFP improves
reliability at moderate overhead for the maximum latencies, despite
less frequently available GTS for sending.

For {\bf Node-initiated} traffic, we compare {\bf Data indication} (a
dedicated indication messages is triggering an Interest by the
consumer) and {\bf Data push}.  In the indication case, we only
consider energy-efficient options in which the Gateway uses the CFP
for Interests.  Our results clearly show an overhead for the three-way
handshake with Data indications. Maximum latencies increase to over
70\,s using the CAP for indication, and to over 240\,s using the CFP,
even in this unstressed scenario. Hence, we ignore Data indication for
the remainder of our evaluation.  In contrast, {\bf Data push} can
obviously be completed in a single message transmission and reduces
the average completion time to 7--11\,s, depending on the CAP/CFP
mapping variant.
{\bf Sending Data in the CAP} is affected by collisions and CSMA
retries. {\bf Sending Data in the CFP} surprisingly reveals a maximum
latency of over 60\,s. We ascribe this to the randomized content
creation interval that leads to occasional synchronized medium access
of several Nodes and then consequently to MAC layer queue processing
delays that last for multiple multi superframes (here 2) with static
slot assignment.

\begin{figure*}
    \centering
    \subfloat[Data unicast (w/ L3 retransmission)]{\includegraphics[width=1.0\columnwidth]{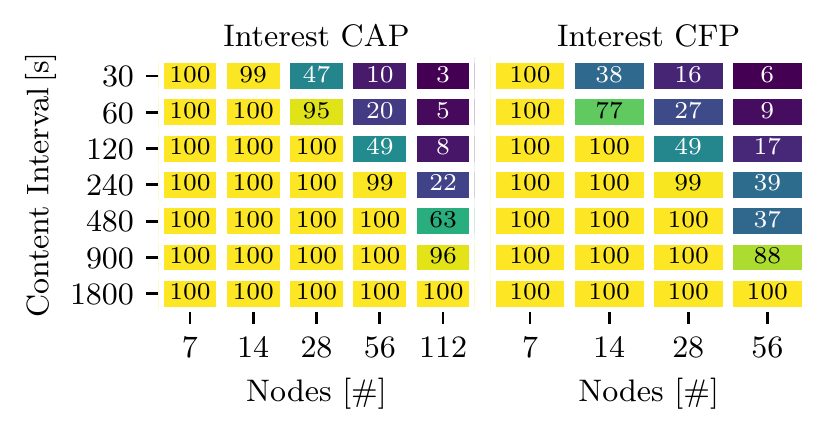}\label{fig:prr_noderequest}}
    \subfloat[Data broadcast (w/o L3 retransmission)]{\includegraphics[width=1.0\columnwidth]{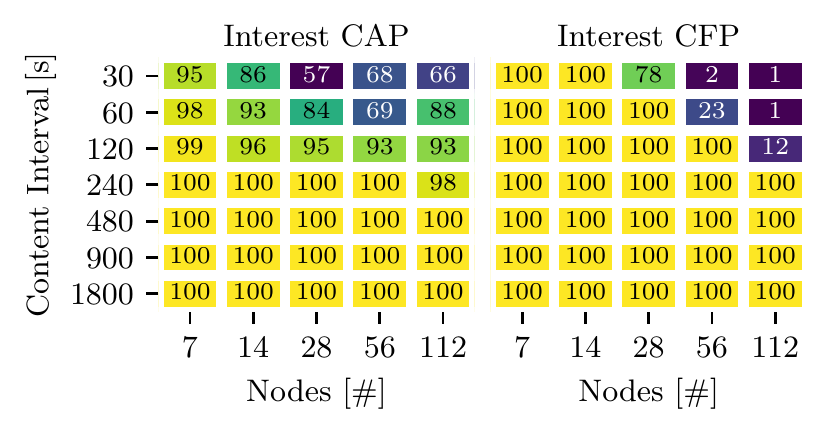}\label{fig:prr_noderequest_bcast}}
    \caption{Success rates [\%] depending on network sizes and content
    invervals for different ICN mappings.}
    \label{fig:success-unicast-broadcast}
\end{figure*}

\autoref{fig:ttc_exponential_4} presents completion times for Data
retrieval from LoRa Nodes in a high data rate scenario (30\,s Data
production intervals, \ref{fig:ttc_30_exponential_4}) and a more
relaxed scenario (900\,s Data production intervals,
\ref{fig:ttc_900_exponential_4}) including ICN retransmissions.  Data
losses result in infinite completion times, hence, the end value of
each graph also reflects its success ratio.

\paragraphS{Performance at high data rates:}
The {\bf Gateway-initiated} requests finish in less than 20\,s, in a
small network of 7 Nodes.  The performance degrades with increasing
networks. In the 14 Nodes case, 90\,\% of the requests are satisfied
in less than 30\,s except for the fully CFP-based mapping.  Missing
Data triggers an Interest retransmission (step at $\approx$ 126\,s)
which reflects our retransmission timeout that aligns with the beacon
interval.  The Interest-CFP/Data-CFP mapping, however, finishes after
600\,s with $\approx$ 50\,\% loss, which is the effect of MAC queue
utilization that has a service rate at the order of one superframe
($\approx 32\,s$).  Network sizes $\geq$ 28 increase the completion
time to the order of hundreds and thousands of seconds at high loss
rates, which then becomes unusable for ICN communication.

Operating in a full CAP mapping exhibits the highest ratio of
 transactions successful at the first attempt  but it inhibits Nodes sleeping. Furthermore, 112 Node
networks reduce the delivery rate to 25\,\% due to collisions and
denied channel access.  Conversely, Interest-CAP/Data-CFP efficiently
combines the ``reactive'' contention-based access for Interests with
reliable contention-free media access for Data.  Interest-CFP/Data-CFP
with 112 Nodes is not possible with our current CFP scheduling
approach (see~\autoref{sec:impl}).

{\bf Node-initiated} Data push reveals faster completion and higher
reliability in comparison to the request-based pattern.
Networks $<$ 28 Nodes show similar behavior for Data-CAP and -CFP
transmissions.  Conversely, networks $\geq$ 28 show the effect of MAC
over-utilization for Data-CFP.  The multi superframe period increases
with the number of Nodes, hence, the average service rate of the MAC
decreases. CAP reduction~\cite{IEEE-802.15.4-16} can mitigate this
effect and will be evaluated in future work.  In contrast to Data-CFP,
Data-CAP performs comparably smooth and transmits $\approx$ 50\,\% of
the messages within 30\,s even under these stressful conditions.

\paragraphS{Performance at low rate:}
In the relaxed scenario (Figure~\ref{fig:ttc_900_exponential_4}),
{\bf Gateway-initiated} requests perform mostly reliable for all mappings
and complete with 100\,\% success after less than 126\,s in networks
$<$ 112 (note the change of the x-axis scale in comparison to Figure~\ref{fig:ttc_30_exponential_4}). For 112 Node networks,
Interest-CAP/Data-CAP clearly shows the effect of CSMA retries.

For {\bf Node-initiated} traffic, Data-CFP is now on-par with Data-CAP
and exhibits the best performance due to the contention-free media
access. In the 112 Node network, the unidirectional push with Data-CAP
faces collisions that cannot be compensated due to the absence of
Interest retransmissions (in contrast to the Gateway-initiated
case). Consequently, Data-CFP is the better option for large networks.

\subsection{Data from Gateway to Node}\label{sec:gw2node}

\paragraphS{Data unicast:} Figure \ref{fig:prr_noderequest} depicts
success rates for Nodes sending Interests to the Gateway with
Interest-CAP and Interest-CFP mappings.  Latencies are less crucial in
this case since Nodes are aware of the constrained regime and sleep
during long round trips. A retransmission will likely be answered by
the Gateway if the requested Data has arrived in the meantime.  Here,
we apply Data-CFP to enable maximum sleep times. Interests are sent in
the CAP or CFP and experience similar challenges as described
in~\autoref{sec:data2gw}. Again, fully CFP-based mappings are not
available for 112 Node networks with the current scheduler.

Figure \ref{fig:prr_noderequest} clearly shows the network operation
boundaries by a diagonal in the heatmap at
30\,s/28\,Nodes--240\,s/112\,Nodes for Interest-CAP, and
30\,s/14\,Nodes--120\,s/240\,Nodes for Interest-CFP. In relaxed
scenarios, both options perform similarly well. Surprisingly,
Interest-CAP outperforms the reliable CFP alternative for larger
networks, despite its best-effort limitation. Hence, Interest
scheduling in a GTS is more susceptible to losses than concurrent
channel access, however, there is a crucial caveat that our
measurements cannot exhibit: In a deployment with multiple Gateways
(in reach), other stub networks share the CAP which increases the
collision probability. In contrast, CFP slots follow a channel hopping
scheme to avoid interfering Nodes. Furthermore, neighboring LoRa
networks can assign the same GTS on orthogonal channels, which
increases the overall throughput. We will focus on multi Gateway
scenarios in future work.

\paragraphS{Data broadcast:} Figure \ref{fig:prr_noderequest_bcast}
depicts success rates for Interests in CAP or CFP and subsequent Data
broadcasts with indirect transmission, triggered by the beacon sent by
the Gateway. Similarly to Interest broadcast
(see~\autoref{sec:data2gw}), the maximum throughput of broadcast Data
packets is limited by the beacon interval. In contrast, however, Data
broadcast can satisfy many pending Interests that have been aggregated
during the beacon period with a single downlink packet. Applying the
ICNLoWPAN encoding, our approach concatenates up to six data items
into one packet. This requires a fixed Interest window size for all
Nodes within one beacon interval and homogeneous content requests during this
period.

Figure~\ref{fig:prr_noderequest_bcast} depicts that the best broadcast
performance can be achieved without Interest retransmissions with
request intervals at the order of 120\,s or greater, which is in line with the
beacon period. The impact of the network size is less significant in
comparison to Figure~\ref{fig:prr_noderequest}, which emphasizes the
advantage of Data broadcast.
Success rates for short request intervals $<$ 120\,s
exhibit the advantage of Interest-CFP over CAP. Exclusive uplink
resources are less susceptible to interference, whereas the shared
media access during CAP suffers from limited media access and
collisions.

Interest retransmissions worsen the success rate for networks $\geq$ 56
Nodes in this scenario, for two reasons: \one delayed request of
``old'' Data occupies the limited downlink resources of the Gateway,
whereas neighbor Nodes simply have to drop duplicate Data. \two
additional transmissions increase media access contention during CAP
and stress the MAC queue during CFP.

\section{LoRa-ICN Convergence Layer}\label{sec:design-conclusions}

Our simulation results revealed that for both upstream
and downstream messages the CFP variants generally provide the best
compromise between low-latency and overall throughput across the wide
range of scenarios we investigated. This leads us to the following
 approach for ICN Interest/Data exchanges. We include the
operations described in~\autoref{sec:design-goals}.

{Interests from  consumers reach  \loraicn producers}  via
the Gateway as per regular ICN forwarding. The Gateway forwards each
Interest that matches a registration in a CFP slot and sets the
expiration timeout to $10 \times n$ seconds, with $n$ set to the
number of registered Nodes. The Gateway should perform Interest
aggregation, \ie suppress duplicate Interests with the
same name. Interests with unknown prefixes are NACKed.
Nodes reply to these Interests with a Data (or a NACK) message in
their assigned CFP slot. This message consumes the Interest on the
Gateway as per regular forwarding behavior. Gateways cache the
content objects in their content store. Depending on the LoRa
network utilization, Interests may expire at the original consumer or
at on-path forwarders. In such cases, consumers should re-issue the
Interest, possibly increasing the Interest expiration time.

{\bf Node Registration} is built on Interests that are transmitted from the Node to the Gateway and adopt NDN prefix registration
\cite{laszz-napp-18}.
  Gateways  propagate their registered
prefixes to adjacent routers, unless scenarios demand otherwise. Registration state needs to be refreshed every 60 minutes. Similar mechanisms could be
used to install per-Node filters to have more fine-granular control
over Interests that are forwarded to the Node.

{\bf Data Provisioning by Nodes} uses unsolicited push (Data messages)
as the primary upstream Data communication primitive, thereby, utilizing a
CFP slot. Gateways will accept unsolicited
Data messages from Nodes that fall under the registered prefix and act
as a custodian, \ie they will keep corresponding Data objects in their
content store and satisfy matching Interests from the Internet. Gateways
should store these objects for several minutes.

For Interests that cannot be satisfied from the content store, the
Gateway performs normal forwarder operations, \ie it forwards the
Interest to matching Nodes following the  prefixes obtained from Node registrations.

{\bf Node Control} actions are triggered by Interests from the
Internet that are intended as a Remote Method Invocation (RMI). They use
the same mechanism as other Interest from Internet consumers (see
above). While this enables basic Node control, it has the usual
problems of using Interest-Data for RMI as described in
\cite{khokp-rrmii-18}.

{\bf Data Retrieval by Nodes} is implemented by Nodes sending
Interests in their CFP slot, setting their local expiration time to
600 seconds. Gateways either consume or forward the
Interest as per regular ICN forwarding. Corresponding Data objects are
cached so that potential Interest retransmission can be satisfied by
the Gateway.

{\bf Downstream Multicast} can be used for synchronized downloads in a
radio resource-efficient way. We assume that this would be triggered
by a control command, possibly referring Nodes to a Manifest pointing
to the actual Data objects. Possible optimization (\eg
Gateway-controlled Data rates) will be studied in future work.

\section{Related Work}\label{sec:related}
\paragraph{ICN and the IoT}\label{sec:related_icniot}
Our work is based on four observations of prior work.
\one the IoT benefits from ICN~\cite{bmhsw-icnie-14}.
\two ICN should not ignore the
MAC~\cite{kgshw-nnmam-17}, to comply with constrained resources.
\three to allow for periodic sleeping of devices without sacrificing
performance, aligning ICN~principles to lower layer frequency-
and time division provides a unique opportunity.  In contrast to prior
work, which presented a design for ICN and 802.15.4 TSCH
mode~\cite{IEEE-802.15.4-16}, we focus on LoRa and DSME.

\paragraph{Analysis of 802.15.4-based Standards}
Comparing TSCH and DSME based on simulations is common~\cite{apmb-saidt-15}.
Choudhury~\etal~\cite{cmml-paimm-20} deploy DSME on constrained Nodes and found that TSCH obtains lower latency and higher throughput for small networks. DSME outperforms TSCH for higher duty cycles and an increasing number of Nodes, though.

Tree-based routing over DSME~\cite{kskt-srasd-20}
has been proposed.
Our topology choice is also supported by the IEEE~802.15 group which
suggests long-range radios operating in star topologies.

\paragraph{Analysis of LoRa and LoRaWAN}
Liando~\etal~\cite{lgtl-kufle-19} provide real-world measurements of
LoRa and LoRaWAN.
Saelens~\etal~\cite{shsp-iedct-18} add
listen-before-talk techniques to overcome band-specific duty cycle
restrictions.
Orfanidis~\etal~\cite{ofjg-cccal-19} find cross-technology
interference between LoRa and 802.15.4 sub-GHz radios and propose an
advanced CCA mechanism for mitigation.
Mikhaylov~\etal~\cite{mfpmm-ealev-19} reveal energy attack vectors in
LoRaWAN, and Shiferaw~\etal~\cite{sak-lcbmc-20} present scalability
issues with LoRaWAN class~B.
All those results
indicate that LoRa and LoRaWAN suffer from scalability issues and are
vulnerable to interference---which motivates our work.

\paragraph{Alternative Protocols for LoRa}
Multi-hop routing in LoRa systems has been analyzed~\cite{ck-lmnrc-20}, including a replacement of the
LoRaWAN MAC by LoRa and IPv6 to make use of RPL implementations for multi-hop networks~\cite{tbs-eticl-17}.
Abrardo~\etal~\cite{aa-mllbm-19} demonstrate a duty-cycling MAC layer to improve sleep time of constrained LoRa devices.
Lee~\etal~\cite{lk-mlisu-18} introduce LoRa mesh-networking that follows a request-response pattern and indicates performance benefits over producer-driven ALOHA.
NDN was deployed on LoRa radios~\cite{lndd-endna-20} which showed the need for a MAC layer.
NDN over WiFi and LoRa~\cite{ks-nrclr-17} was proposed to connect `isolated regions', however, nothing was mentioned about the LoRa MAC and how it prevents wireless interference and energy depletion.

For contention-based MAC, experiments with CSMA and CAD in LoRa-type
networks indicate performance gains \cite{p-iecca-18},
without providing specific LoRa measurement results, though.
Logical channel LoRa PHY configurations may assist frequency- and time
division multiple access protocols \cite{gbv-slpld-18}.
Listen-before-talk in sub-GHz
bands~\cite{lbbp-calma-20}  performs better than ALOHA in LoRaWAN, and unconfirmed messages
perform better in dense deployments.

Adaptations for time-slotted~\cite{zakp-ttlii-20} LoRa have been presented in~\cite{rffsg-uliwn-17,hoof-tlrri-20}
but consist of only three Nodes and limited traffic (12~packet/h).
Alamos~\etal~\cite{aksw-edmls-22} introduce DSME-LoRa and deploy 15 nodes.
In this paper, we close the gap by analysing reliability in larger networks.

\section{Conclusions and Outlook}\label{sec:outlook}

The LoRa PHY is a radio layer that caters to many long-range,
low-power communication scenarios. Unfortunately, the commonly used
upper layer, LoRaWAN, is a  vertically integrated
communication system that cannot provide direct Internet connectivity
and direct data sharing, but leads to centralized system
architectures of limited scalability.

We introduced \loraicn to overcome these limitations.
We designed a new LoRa system from the ground up, leveraging the
existing LoRa PHY but employing IEEE 802.15.4 DSME as a MAC layer and
ICN as a network layer. To that end, we have defined a suitable DSME
configuration, specified mappings of ICN protocol messages to DSME
mechanisms, and proposed specific ICN extensions and Node
requirements.  Our DSME implementation provides the benefits of
horizontal scalability, deterministic media access, and low-power
operations.
We could show in simulations  for common network sizes that ICN messages gain reliability and reduce  latency   when mapped to DSME-CFP messages.

To support the current most relevant use case of IoT data
transmission from constrained devices to the Internet well, we added a
data custodian feature to Gateways.
The result is a new LoRa system that supports direct end-to-end
communication with LoRa Nodes and that can provide additional features
such as downstream multicast natively.
We claim that this highlights the versatility of ICN as an IoT network
layer: By leveraging and minimally extending standard ICN caching, a
LoRa Gateway can connect a delay-prone LoRa network to the Internet,
without requiring any application awareness or protocol translation.

Our future work includes investigating other LoRa configurations (mesh
networks), large-scale operation, ICN security in LoRa,
and using RICE~\cite{khokp-rrmii-18} for enhanced Node control.

\paragraph{Artifacts}
Our code is available on
\url{https://github.com/inetrg/IFIP-Networking-LoRa-ICN-2022}

\label{lastbodypage}

\balance
\bibliographystyle{IEEEtran}
\bibliography{}

\label{lastpage}

\end{document}